%% file: main.tex
\documentclass[conference]{IEEEtran}
\IEEEoverridecommandlockouts
\usepackage{cite}
\usepackage{amsmath,amssymb,amsfonts}
\usepackage{algorithmic}
\usepackage{graphicx}
\usepackage{textcomp}
\usepackage{xcolor}
\usepackage{soul}
\usepackage[labelfont=bf]{caption}
\usepackage{booktabs}
\usepackage{graphicx}          
\usepackage{tikz}              
\usetikzlibrary{positioning,arrows.meta,shapes}

\begin{document}

\title{LLM-Driven Adaptive 6G-Ready Wireless Body Area Networks: Survey and Framework}

\author{
    \IEEEauthorblockN{
        Mohammad Jalili Torkamani\IEEEauthorrefmark{1},
        Negin Mahmoudi\IEEEauthorrefmark{2},
        Kiana Kiashemshaki\IEEEauthorrefmark{3}
    }
    \IEEEauthorblockA{\IEEEauthorrefmark{1}School of Computing, University of Nebraska--Lincoln, Lincoln, NE, USA\\
    mJaliliTorkamani2@huskers.unl.edu}
    \IEEEauthorblockA{\IEEEauthorrefmark{2}Department of Civil, Environmental, and Ocean Engineering, Stevens Institute of Technology, Hoboken, NJ, USA\\
    Email: nmahmoud1@stevens.edu}
    \IEEEauthorblockA{\IEEEauthorrefmark{3}Department of Computer Science, Bowling Green State University, Bowling Green, OH, USA\\
    Email: kkiana@bgsu.edu}
}

\maketitle

\begin{abstract}
Wireless Body Area Networks (WBANs) enable continuous monitoring of physiological signals for applications ranging from chronic disease management to emergency response. Recent advances in 6G communications, post-quantum cryptography, and energy harvesting have the potential to enhance WBAN performance. However, integrating these technologies into a unified, adaptive system remains a challenge. This paper surveys some of the most well-known Wireless Body Area Network (WBAN) architectures, routing strategies, and security mechanisms, identifying key gaps in adaptability, energy efficiency, and quantum-resistant security. We propose a novel Large Language Model-driven adaptive WBAN framework in which a Large Language Model acts as a cognitive control plane, coordinating routing, physical layer selection, micro-energy harvesting, and post-quantum security in real time. Our review highlights the limitations of current heuristic-based designs and outlines a research agenda for resource-constrained, 6G-ready medical systems. This approach aims to enable ultra-reliable, secure, and self-optimizing WBANs for next-generation mobile health applications.
\end{abstract}


\input{introduction}
\input{architecture}
\input{routing_protocols_in_wban}

\input{challenges_and_solutions}
\input{proposed_framework}
\input{future_challenges_direction}
\input{conclusion}

\bibliographystyle{IEEEtran}
\bibliography{list}

\end{document}

%% file: introduction.tex
\section{Introduction}
\label{sec:introduction}

Wireless Body Area Networks (WBANs) are low-power, short-range communication systems designed to continuously monitor vital signs and physiological parameters for a variety of health-related applications \cite{cornet2022overview}. As mentioned by Cornet et al. \cite{cornet2022overview}, WBAN nodes can be worn on the skin, implanted in the body, or carried externally, and they transmit data wirelessly to a coordinator, which forwards information to remote servers or edge devices for analysis. This architecture enables continuous health monitoring, real-time alerts, and intelligent diagnostic support. 

Applications of WBANs span diverse domains: from elderly care and chronic disease management to high-risk occupational safety (e.g., firefighters, soldiers) \cite{zhang2014ubiquitous}, assistive services for people with disabilities, and fitness tracking. Even in populations without immediate health risks, WBANs facilitate early disease detection, emergency response, and personalized lifestyle optimization.

Despite their promise, WBAN deployment faces multiple challenges: 
\begin{itemize}
    \item \textbf{Energy Efficiency:} WBAN nodes operate under strict power constraints, making energy harvesting, duty cycling, and transmission optimization essential.
    \item \textbf{Security and Privacy:} Sensitive medical data requires robust protection against cyberattacks, including future quantum-capable threats \cite{jian2024systematic}.
    \item \textbf{Reliability and Latency:} Continuous monitoring demands near-real-time data transfer, even under high mobility and variable channel conditions.
    \item \textbf{Biocompatibility and Safety:} Devices must minimize tissue heating, avoid harmful emissions, and maintain long-term physiological compatibility.
\end{itemize}

The ongoing transition from 5G to 6G networks opens new opportunities for WBAN performance enhancement through ultra-reliable low-latency communications (URLLC), terahertz (THz) links, and intelligent edge integration \cite{selem2019ehealth}. Similarly, advances in post-quantum cryptography (PQC) promise to strengthen long-term security for medical data. Meanwhile, recent breakthroughs in Large Language Models (LLMs) suggest a new paradigm for intelligent, context-aware WBAN control: LLMs can act as a cognitive control plane, dynamically optimizing routing, link selection, and security parameters based on multi-modal sensor and network telemetry. LLMs have shown strong potential for adaptive reasoning and decision-making in complex domains such as software engineering \cite{lee2024llm, torkamani2025streamlining, torkamani2024assertify, kiashemshaki2025secure}, suggesting their applicability to dynamic WBAN management.
Similar AI-driven frameworks have demonstrated high performance in other critical healthcare applications, including brain tumor classification from MRI scans~\cite{golkarieh2025breakthroughs} and automated lung cancer detection with advanced segmentation models~\cite{golkarieh2025advanced}.
and early disease detection via wearable sensor networks~\cite{muthu2020iot, sensorNet2024}.

Despite these technological advances, current WBAN designs remain fragmented: routing, security, and energy management are often addressed in isolation, relying on static rules rather than adaptive, context-aware intelligence. This gap motivates our integrated, LLM-driven approach.

\subsection{Paper Contributions}
In this paper, we make the following contributions:
\begin{itemize}
    \item \textbf{Survey:} We provide a review of WBAN architectures, routing strategies, and security mechanisms, and emphasize trade-offs between energy efficiency, reliability, and security.

    \item \textbf{Gap Analysis:} We identify research gaps in current WBAN solutions, particularly in adaptive backhaul selection, physiological safety-aware routing, and quantum-safe security integration.
    \item \textbf{Conceptual Framework:} We introduce an LLM-driven adaptive WBAN control framework, which combines multi-layer adaptability, context-aware decision-making, micro-energy harvesting, and PQC-compliant security.
    \item \textbf{Research Direction:} We outline future research directions for integrating LLM reasoning, 6G physical layers, and edge-based AI in WBANs.
\end{itemize}

\subsection{Paper Organization}
The remainder of this paper is structured as follows. 
Section~\ref{sec:architecture} provides an overview of WBAN architectures and communication models. 
Section~\ref{sec:routing} reviews routing strategies and analyzes associated performance trade-offs. 
Section~\ref{sec:challenges_and_solutions} examines WBAN security challenges and highlights recent solution approaches. 
Section~\ref{sec:framework} introduces the proposed LLM-driven adaptive WBAN framework. 
Section~\ref{sec:future_challenges_direction_conclusion} discusses future research directions, and 
Section~\ref{sec:conclusion} summarizes the main contributions of the paper.

%% file: architecture.tex
\section{Architecture}
\label{sec:architecture}

The IEEE 802.15.6 standard defines Wireless Body Area Network (WBAN) communications in multiple scenarios based on device placement and communication medium \cite{kwak2010overview}. For clarity, we categorize WBAN components and communication links into two primary domains: \emph{internal} and \emph{external}.

\begin{itemize}
    \item  \textbf{Internal Devices and Human Body Communication (HBC):}
    Internal WBAN devices are implanted or ingested sensors designed to interface directly with human organs or tissues. These devices must be engineered for biocompatibility, durability, and minimal physiological disruption. Environmental challenges include variable temperature, chemical exposure, and the inability to easily replace or recharge devices once implanted. 

Communication between internal nodes often employs \emph{Human Body Communication} (HBC), first introduced by Zimmerman in 1996 \cite{zimmerman1996personal}. In HBC, the human body serves as a conductive channel, typically operating in the 5–100~MHz range with bandwidths of 3–5~MHz. While HBC offers low radiation exposure and reduced power consumption compared to traditional radio frequency (RF) links, it suffers from:
\begin{itemize}
    \item High and variable path loss influenced by anatomy, posture, and sensor placement \cite{hwang2014empirical}.
    \item Limited data rates compared to millimeter-wave links \cite{7481217}.
    \item Sensitivity to external interference in shared frequency ranges \cite{9436010}.
\end{itemize}

    \item  \textbf{External Devices and RF-Based Communication:}
    External devices include wearables (e.g., smartwatches, headbands) and body-mounted nodes, which are easier to access and maintain compared to implanted devices. These nodes often serve as coordinators or relays, bridging internal sensors with external infrastructure. 

Historically, WBANs have relied on the 2.4~GHz Industrial, Scientific, and Medical (ISM) band, but this spectrum is increasingly congested. Recent studies have explored the 60~GHz millimeter-wave (mmWave) band for short-range, high-throughput WBAN communications \cite{akbar2022wireless}. The mmWave band supports multi-gigabit rates and low latency but faces:
\begin{itemize}
    \item High propagation loss, particularly in non-line-of-sight (NLOS) conditions.
    \item Increased susceptibility to blockage and shadowing by body parts.
    \item Complex beamforming requirements to maintain stable links \cite{selem2019ehealth}.
\end{itemize}

Emerging 6G concepts such as \emph{Terahertz (THz) communication}, \emph{Optical Wireless Technologies (OWT)}, and \emph{Intelligent Reflecting Surfaces (IRS)} are expected to further extend WBAN external communication capabilities \cite{selem2019ehealth}. These technologies promise ultra-high data rates, interference mitigation, and improved energy efficiency.\\*
    
\end{itemize}

To address the constraints of internal and external WBAN links, future architectures must dynamically adapt link selection, resource allocation, and data flow, which sets the stage for hybrid and LLM-driven approaches.

\subsection{Architectural Considerations and Hybrid Designs}
A major challenge in WBAN architecture is the \emph{non-homogeneous relay environment}. Communication paths often combine heterogeneous links (e.g., HBC for implants, mmWave for on-body links, LoRaWAN or Wi-Fi for backhaul). The slowest or least reliable link becomes the bottleneck for end-to-end performance. 

Hybrid designs increasingly leverage:
\begin{itemize}
    \item \textbf{Adaptive Link Selection:} Switching between HBC, RF, or optical links based on instantaneous link quality and application requirements.
    \item \textbf{Energy-Aware Relay Placement:} Positioning intermediate nodes to minimize transmission power and extend battery life \cite{yaghoubi2022wireless}.
    \item \textbf{Context-Aware Routing:} Dynamically selecting routes based on posture, mobility, or environmental interference \cite{akbar2022wireless}.
\end{itemize}

\subsection{LLM-Driven Architectural Evolution}
While current adaptive architectures rely on pre-defined heuristics, we envision the use of LLMs as cognitive control agents within WBAN coordinators or edge gateways. Integrating multi-modal sensor data for real-time decision-making has already shown clinical impact in IoT-based remote health monitoring systems~\cite{IoThealth2023}, supporting the feasibility of LLM-based reasoning in WBAN architectures.
By continuously ingesting multi-modal telemetry (e.g., link quality indicators, battery levels, physiological metrics), an LLM could:
\begin{itemize}
    \item Infer optimal PHY/backhaul combinations in real time.
    \item Anticipate link degradation due to posture or environmental changes.
    \item Balance energy harvesting strategies with quality-of-service (QoS) requirements.
    \item Recommend secure key management updates aligned with post-quantum cryptography standards.
\end{itemize}
This paradigm shifts WBAN control from reactive, rule-based switching to proactive, AI-driven optimization, paving the way for self-configuring health monitoring networks in the 6G era.

%% file: routing_protocols_in_wban.tex
\section{Routing Protocols in WBAN}
\label{sec:routing}

Routing in Wireless Body Area Networks (WBANs) is uniquely challenging due to the combination of strict energy constraints, constantly changing topologies caused by human movement, and safety requirements such as limiting tissue heating \cite{akbar2022wireless,qu2019survey}. Unlike conventional Wireless Sensor Networks (WSNs), WBANs must deliver high reliability and low latency for medical data streams, often in environments with unpredictable signal conditions.

\subsection{Taxonomy of Routing Approaches}
According to the recent surveys \cite{cornet2022overview,yaghoubi2022wireless,narwal2024dissecting}, WBAN routing strategies can be broadly categorized into the following families:

\begin{enumerate}
    \item \textbf{Temperature-Aware Routing}  
    Avoids overheated nodes to reduce the risk of tissue heating and prevent performance degradation. While this approach improves safety, it can increase delay by forcing packets to take longer paths.

    \item \textbf{Posture- and Mobility-Based Routing}  
    Uses posture detection (e.g., standing, walking, lying) to anticipate link breaks caused by body movement \cite{qu2019survey}. This predictive capability improves reliability but requires additional sensing and processing.

    \item \textbf{Cluster-Based Routing}  
    Groups nodes into clusters, with a rotating cluster head handling inter-cluster communication \cite{akbar2022wireless}. This balances energy consumption across nodes but introduces coordination overhead.

    \item \textbf{QoS-Aware Routing}  
    Prioritizes traffic based on application requirements—such as real-time ECG streaming versus periodic glucose monitoring—by optimizing for latency, throughput, or reliability \cite{yaghoubi2022wireless}.

    \item \textbf{Cross-Layer Routing}  
    Coordinates decision-making across the PHY, MAC, and network layers, for example by using physical layer link quality metrics to guide routing decisions.
\end{enumerate}

These categories are not mutually exclusive; hybrid approaches are common. For example, a protocol might combine QoS prioritization with temperature constraints, or mobility prediction with cluster-based organization.

\subsection{Performance Trade-Offs}
Different routing approaches make distinct trade-offs between latency, energy efficiency, and network robustness:
\begin{itemize}
    \item Low-latency methods, such as posture-based or QoS-prioritized routing, tend to consume more energy due to frequent route recalculations and control message exchanges.
    \item Energy-efficient designs, such as temperature-aware or cluster-based routing, often sacrifice latency, making them unsuitable for urgent alerts.
    \item Load-balancing schemes, such as multipath or rotating cluster head designs, can extend network lifetime but add coordination overhead.
\end{itemize}
The suitability of a routing protocol depends heavily on the application context—what works well for static patients in clinical environments may not be optimal for mobile users in outdoor settings.

\subsection{Emerging Trends}
Recent research points to several promising trends:
\begin{enumerate}
    \item \emph{Context-Aware Routing:} Protocols that dynamically adapt routes based on posture, mobility, and link quality in real time.
    \item \emph{Machine Learning-Assisted Routing:} Leveraging supervised and reinforcement learning to predict link quality and select optimal paths.
    \item \emph{Edge-Assisted Routing:} Offloading complex routing decisions to nearby edge nodes to reduce the processing burden on WBAN devices.
\end{enumerate}

\subsection{LLM-Driven Adaptive Routing Concept}
Building on these trends, we envision an \emph{LLM-driven adaptive routing framework} in which a Large Language Model acts as a high-level decision engine for WBANs. The LLM would:
\begin{itemize}
    \item Interpret multi-modal telemetry, including link quality indicators, motion patterns, energy status, and physiological
    constraints.
    \item Predict potential link degradation or node failure before it occurs.
    \item Recommend routing mode changes—such as switching between QoS-prioritized, temperature-aware, or cluster-based configurations—based on anticipated conditions.
    \item Coordinate routing decisions with security requirements, ensuring that critical data flows traverse secure, reliable links.
\end{itemize}
This approach shifts routing from a static, rule-based process to a flexible, predictive system that can proactively adapt to the dynamic and safety-critical nature of WBAN operations.
Machine learning-assisted routing strategies for WBANs, such as reinforcement learning and graph-based optimization~\cite{MLrouting2024}, provide a precedent for the predictive adaptation we envision with LLM control.

%% file: challenges_and_solutions.tex
\section{WBAN Security Challenges and Recent Solutions}
\label{sec:challenges_and_solutions}

Security and privacy are critical for Wireless Body Area Networks (WBANs), as they handle highly sensitive medical information. Unlike traditional wireless sensor networks, WBANs must balance strong security requirements with stringent energy and latency constraints \cite{jian2024systematic}. The close integration with the human body also imposes unique constraints: excessive computation or communication for security purposes can increase heat generation or deplete limited power resources, potentially compromising patient safety.

\subsection{Threat Landscape}
Recent surveys \cite{yaghoubi2022wireless,jian2024systematic} categorize WBAN security threats into several classes:
\begin{enumerate}
    \item \textbf{Eavesdropping and Traffic Analysis}: Interception of physiological data in transit, which can reveal sensitive health conditions.
    \item \textbf{Data Tampering and Injection}: Alteration of sensed data or injection of falsified readings, potentially leading to harmful medical decisions.
    \item \textbf{Replay Attacks}: Reuse of captured valid data packets to mislead monitoring systems.
    \item \textbf{Denial-of-Service (DoS)}: Disruption of network availability through jamming, flooding, or protocol exploitation.
    \item \textbf{Key Compromise and Cloning}: Extraction or duplication of cryptographic keys, undermining authentication and confidentiality.
\end{enumerate}

\subsection{Constraints on Security Mechanism Design}
Security protocols for WBANs must adhere to several strict constraints:
\begin{itemize}
    \item \textbf{Low Energy Overhead:} Cryptographic \cite{joseph2022transitioning, kiashemshaki2025securescalableblockchainvoting} operations must not significantly reduce network lifetime.
    \item \textbf{Low Latency:} Emergency medical alerts require rapid authentication and transmission.
    \item \textbf{Physiological Safety:} Prolonged computation or communication must not cause unsafe thermal effects.
    \item \textbf{Interoperability:} Protocols should remain compatible with IEEE 802.15.6 and other healthcare communication standards.
\end{itemize}

\subsection{Current Security Approaches}
\paragraph{Lightweight Cryptography}  
Block ciphers such as AES-128 and AES-256, combined with lightweight hash functions (e.g., SHA-256 variants), remain common due to their standardized security guarantees and relatively low computational footprint \cite{akbar2022wireless}.

\paragraph{Elliptic Curve Cryptography (ECC)}  
ECC-based schemes provide strong security with shorter key sizes than RSA, reducing computation and communication overhead \cite{yaghoubi2022wireless}. However, ECC is still vulnerable to quantum attacks.

\paragraph{Post-Quantum Cryptography (PQC)}  
Recent work has begun integrating lattice-based Key Encapsulation Mechanisms (KEMs) and code-based signatures into WBANs to prepare for quantum-capable adversaries \cite{jian2024systematic}. These algorithms are designed for low-power devices, but widespread adoption remains limited due to implementation complexity.

\paragraph{Biometric-Based Authentication}  
Physiological signals such as ECG or PPG patterns can serve as user-specific cryptographic keys, reducing reliance on stored keys. While promising, these methods require careful calibration to avoid false rejection or acceptance rates.

\subsection{Open Security Challenges}
The current landscape of WBAN security still faces the following challenges:
\begin{itemize}
    \item \emph{Balancing Security and Energy Efficiency:} Most PQC schemes require more computation than classical algorithms, necessitating intelligent adaptation.
    \item \emph{Dynamic Key Management:} Key update frequency and distribution strategies must account for node mobility, topology changes, and intermittent connectivity.
    \item \emph{Threat Prediction:} Existing systems react to detected threats but rarely anticipate them.
\end{itemize}

\subsection{LLM-Enhanced Security Management}
We propose incorporating LLMs into WBAN security management as part of an intelligent control plane. In this role, the LLM would:
\begin{itemize}
    \item Continuously analyze security logs, network traffic patterns, and environmental context to detect anomalies.
    \item Predict likely attack vectors (e.g., replay or jamming) before they disrupt the network.
    \item Recommend cryptographic parameter adjustments, such as switching from lightweight AES-128 to PQC-based KEMs when a high-risk state is detected.
    \item Coordinate with routing and backhaul selection to ensure that critical data flows are both secure and efficient.
\end{itemize}

This AI-driven approach shifts security from static, preconfigured policies toward proactive, context-aware defense strategies that can adapt to evolving threats while preserving WBAN performance and patient safety.

The architectural flexibility, routing adaptability, and security resilience discussed above highlight the need for a unifying control mechanism. In the next section, we synthesize these perspectives into a single LLM-driven adaptive WBAN framework capable of orchestrating decisions across layers in real time.

%% file: proposed_framework.tex
\section{Proposed LLM-Driven Adaptive WBAN Framework}
\label{sec:framework}

Drawing on the architectural, routing, and security insights reviewed in previous sections, we propose a conceptual \emph{LLM-driven adaptive WBAN framework} for 6G-ready, energy-aware, and secure mobile health systems. Unlike conventional heuristic-driven designs, this framework leverages an LLM as a cognitive control plane to coordinate communication, energy, and security decisions across the network.

\subsection{Framework Overview}
The framework integrates four primary capabilities:
\begin{enumerate}
    \item \textbf{Adaptive PHY and Backhaul Selection}: The system dynamically chooses between Human Body Communication (HBC), mmWave, Terahertz, Optical Wireless Technologies (OWT), or LoRaWAN backhaul links based on instantaneous link quality, per-bit energy cost, and latency requirements.
    
    \item \textbf{LLM-Assisted Context-Aware Routing}: A centralized LLM processes real-time telemetry, including link quality, posture and motion data, residual battery levels, and application QoS requirements, to select optimal routing strategies (e.g., QoS-prioritized, temperature-aware, or cluster-based) and reconfigure paths proactively.
    
    \item \textbf{Micro–Energy Harvesting Integration}: The LLM incorporates models of each node’s harvesting capability (e.g., thermal, piezoelectric, solar) to anticipate future energy availability and adjust duty cycling, routing, and transmission scheduling accordingly.
    
    \item \textbf{Post-Quantum-Safe Security Orchestration}: Security management adapts in real time, allowing the LLM to recommend cryptographic transitions (e.g., from AES to a PQC-based KEM) when elevated risk is detected, while maintaining lightweight operations under normal conditions.
\end{enumerate}

\subsection{Role of the LLM in Control and Optimization}
The LLM serves as a reasoning layer capable of:
\begin{itemize}
    \item \emph{Multi-modal Input Processing:} Interpreting heterogeneous data streams, from link-layer statistics to physiological sensor readings, using structured prompts\cite{Yin_2024}.
    \item \emph{Predictive Decision-Making:} Anticipating link degradation, energy depletion, or potential security breaches based on historical and contextual patterns \cite{wei2025plangenllmsmodernsurveyllm}.
    \item \emph{Coordinated Policy Recommendations:} Proposing synchronized adjustments to PHY selection, routing mode, and security protocols to maintain optimal service  \cite{jiang2024linkslargelanguagemodel}.
    \item \emph{Knowledge Refinement:} Updating its internal models based on feedback from past decisions to improve future recommendations \cite{madaan2023selfrefineiterativerefinementselffeedback}.
\end{itemize}

\subsection{Operational Workflow}
The framework operates in a cyclical process:
\begin{enumerate}
    \item \textbf{Data Collection:} WBAN nodes and coordinators periodically measure network conditions, energy levels, and relevant physiological metrics.
    \item \textbf{Inference:} Collected data is provided to the LLM, which interprets the state of the network and predicts near-future conditions.
    \item \textbf{Recommendation:} The LLM produces prioritized reconfiguration actions for routing, link selection, and security parameters.
    \item \textbf{Validation and Execution:} Lightweight verification ensures safety, latency, and energy constraints are respected before implementing changes.
    \item \textbf{Feedback:} Post-deployment performance metrics are logged and used to refine future decisions.
\end{enumerate}

\subsection{Advantages Over Traditional Approaches}
Compared with existing WBAN designs, the proposed framework offers:
\begin{itemize}
    \item \emph{Proactive Adaptation:} Shifts from reactive management to predictive, preemptive adjustments.
    \item \emph{Holistic Optimization:} Simultaneously considers performance, safety, energy, and security factors in decision-making.
    \item \emph{Post-Quantum Readiness:} Incorporates quantum-resistant security as a dynamic, context-dependent layer.
    \item \emph{Incremental Deployability:} Allows partial adoption, such as introducing LLM-based routing first, then extending to security and energy management.
\end{itemize}

%% file: future_challenges_direction.tex
\section{Future Research Directions}
\label{sec:future_challenges_direction_conclusion}

The proposed LLM-driven adaptive WBAN framework opens multiple avenues for exploration at the intersection of wireless communication, artificial intelligence, and healthcare technology. Realizing its potential will require advances in both enabling technologies and their integration into medical ecosystems.

\subsection{LLM Adaptation for Resource-Constrained Environments}
We suggest the future work of this research to focus on:
\begin{itemize}
    \item \textbf{Model Compression and Distillation:} Creating smaller, specialized LLMs for low-power edge devices.
    \item \textbf{On-Device Inference Optimization:} Applying quantization, pruning, and low-rank adaptation to reduce latency and energy use.
    \item \textbf{Federated Fine-Tuning:} Updating models locally without transmitting sensitive data.
\end{itemize}

\subsection{Integration with 6G and Emerging Physical Layers}
We believe that involving such LLM-driven solutions will facilitate:
\begin{itemize}
    \item Leveraging terahertz bands for high-capacity medical data transfer.
    \item Deploying intelligent reflecting surfaces to enhance link reliability.
    \item Hybrid PHY switching under LLM control for dynamic QoS-energy balancing.
\end{itemize}

\subsection{Post-Quantum and Biometric Security}
In terms of post-quantum and biometric security, we suggest:
\begin{itemize}
    \item Combining physiological and behavioral biometrics for continuous authentication.
    \item AI-driven adaptive security policies based on real-time threat assessment.
\end{itemize}

\subsection{LLM-Enabled Predictive Healthcare}
Beyond network optimization, LLMs could:
\begin{itemize}
    \item Analyze longitudinal sensor data for early anomaly detection.
    \item Generate personalized health insights and alerts.
    \item Integrate seamlessly with electronic health records for physician decision support.
\end{itemize}

\subsection{Ethics, Privacy, and Regulation}
In terms of the ethical considerations, deployment must ensure:
\begin{itemize}
    \item Data sovereignty for patients and healthcare providers.
    \item Transparent and explainable LLM decision-making.
    \item Compliance with HIPAA, GDPR, and medical device certification standards.
\end{itemize}

\subsection*{\textbf{Key Takeaways}}
We believe that the future research in this domain should explore the following directions:
\begin{enumerate}
    \item \textbf{Energy-efficient LLM adaptation}: Develop optimization techniques that enable LLMs to operate within the computational and power constraints of embedded WBAN devices without sacrificing inference accuracy.
    
    \item \textbf{Seamless integration with 6G and hybrid communication layers}: Design architectures that allow WBANs to interoperate smoothly with emerging 6G infrastructures and heterogeneous link technologies and ensure continuous connectivity and minimal latency.

    \item \textbf{Quantum-resistant, multi-factor security}: Implement cryptographic protocols that can withstand quantum computing attacks, complemented by layered authentication mechanisms tailored to the resource limitations and safety requirements of WBAN environments.
        
    \item \textbf{Clinically interpretable AI}: Develop AI models that make their predictions and reasoning processes transparent to healthcare professionals and enable real-time clinical validation and actionable decision support without requiring specialist knowledge in AI.

    \item \textbf{Regulatory and ethical alignment}: Establish design, deployment, and operational practices that adhere to healthcare regulations, protect patient privacy, and address ethical concerns such as informed consent, data ownership, and bias mitigation.

\end{enumerate}

%% file: conclusion.tex
\section{Conclusion}
\label{sec:conclusion}

This paper has presented a structured survey of recent advancements in Wireless Body Area Networks (WBANs), covering architectures, routing strategies, and security mechanisms with a focus on developments from 2020 to 2025. Our review highlights persistent challenges in achieving energy efficiency, low latency, reliability, and robust security—particularly in the context of emerging 6G communication paradigms and the impending need for post-quantum cryptographic readiness. Building on these insights, we have proposed a conceptual \emph{LLM-driven adaptive WBAN framework} that unifies four critical capabilities: adaptive PHY/backhaul selection, context-aware routing, micro–energy harvesting integration, and dynamic, post-quantum-safe security orchestration. At the core of this architecture is a Large Language Model (LLM) acting as a cognitive control plane, capable of interpreting multi-modal telemetry, predicting performance or security risks, and recommending real-time reconfigurations that balance quality-of-service, safety, and energy constraints. The proposed framework shifts WBAN operation from static, heuristic-based management toward proactive, knowledge-rich, and self-optimizing control. By leveraging LLM reasoning in conjunction with emerging 6G physical layers and quantum-resilient security, the architecture lays the groundwork for ultra-reliable, intelligent, and secure mobile health ecosystems. Future work should address practical deployment challenges, including the adaptation of LLMs for resource-constrained environments, ensuring explainability in clinical contexts, and meeting regulatory compliance for medical device certification. The convergence of AI-driven decision-making, advanced wireless communication, and secure healthcare data management holds the potential to significantly enhance patient monitoring, preventive medicine, and emergency response in next-generation WBAN deployments. In summary, the integration of LLM intelligence into WBAN design represents a promising paradigm shift—one that can transform mobile health systems into highly adaptive, context-aware, and resilient networks, capable of meeting the demands of an increasingly connected and data-driven healthcare landscape.